\documentclass[aps,prd,preprint,nofootinbib,showpacs,showkeys,preprintnumbers,amsmath,amssymb,11pt]{revtex4}

\usepackage{graphicx}% Include figure files
\usepackage{dcolumn}% Align table columns on decimal point
\usepackage{bm}% bold math
\usepackage{bbm}

\begin{document}

%\preprint{}

\title{Search for Sub-eV Sterile Neutrinos in the Precision Multiple Baselines Reactor Antineutrino Oscillation Experiments\vspace{0.3cm}}

\author{\bf Shu Luo}
\email{luoshu@xmu.edu.cn}                                                                                                                                                                                                                                                                                                                                                                                                                                                                                                                                                                                                                                                                                                                                                                                                                                                                                                                                                                                                                                                                                                                                                                                                                                                                                                                                                                                                                                                                                                                                                                                                                                                                                                                                                                                                                                                                                                                                                                                                                                                                                                                                                                                                                                                                                                                                                                                                                                                              	

\affiliation{Department of Astronomy and Institute of Theoretical Physics and Astrophysics, Xiamen University, Xiamen, Fujian 361005, China\vspace{0.9cm}}

\begin{abstract}
According to different effects on neutrino oscillations, the unitarity violation in the MNSP matrix can be classified into the {\it direct} unitarity violation and the {\it indirect} unitarity violation which are induced by the existence of the light and the heavy sterile neutrinos respectively. Of which sub-eV sterile neutrinos are of most interesting. We study in this paper the possibility of searching for sub-eV sterile neutrinos in the precision reactor antineutrino oscillation experiments with three different baselines at around 500 m, 2 km and 60 km. We find that the antineutrino survival probabilities obtained in the reactor experiments are sensitive only to the direct unitarity violation and offer very concentrated sensitivity to the two parameters $\theta^{}_{14}$ and $\Delta m^{2}_{41}$. If such light sterile neutrinos do exist, the active-sterile mixing angle $\theta^{}_{14}$ could be acquired by the combined rate analysis at all the three baselines and the mass-squared difference $\Delta m^{2}_{41}$ could be obtained by taking the Fourier transformation to the $L / E$ spectrum. Of course, for such measurements to succeed, both high energy resolution and large statistics are essentially important. 
\end{abstract}



\maketitle

\section{Direct and Indirect Unitarity Violation in the Lepton Flavor Mixing Matrix}

Besides the three known active neutrinos $\nu^{}_e$, $\nu^{}_\mu$ and $\nu^{}_\tau$, there  may exist additional sterile neutrinos which do not directly take part in the weak interactions except those induced by the mixing with active neutrinos \cite{sterile}. In the presence of $n$ generations of sterile neutrinos, the $3 \times 3$ Maki-Nakagawa-Sakata-Pontecorvo (MNSP) matrix \cite{MNSP} is the submatrix of the full $(3+n) \times (3+n)$ unitary mixing matrix. If there is small mixing between the active and the sterile neutrinos, the MNSP matrix must be slightly non-unitary. According to the different effects on neutrino oscillations, the unitarity violation in the MNSP matrix can be classified into two categories: {\it direct} unitarity violation and {\it indirect} unitarity violation \cite{Xing2013}.
\begin{itemize}
\item The indirect unitarity violation is brought by the existence of heavy sterile neutrinos, which themselves are too massive to be kinematically produced in the neutrino oscillation experiments. The heavy right-handed sterile neutrinos are natural ingredients of the canonical type-I seesaw mechanism \cite{typeI} and some other seesaw models \cite{seesaw}.
\item The direct unitarity violation is caused by the existence of light sterile neutrinos which are able to participate in neutrino oscillations as their active partners. The sterile neutrinos with masses $m \sim {\cal O}(1)$ eV are proposed to explain the LSND \cite{LSND}, MiniBooNE \cite{MiniBooNE}, reactor antineutrino \cite{reactor} and Gallium \cite{Gallium} anomalies. Furthermore, current cosmological observations \cite{Cosmology} still allow the existence of sub-eV sterile neutrinos. 
\end{itemize}

To study their different effects on neutrino oscillations, we consider in a special ($3$+$\mathbbm{1}$+$\mathbf{1}$) framework where $\mathbbm{1}$ light sterile neutrino $\nu^{}_{s}$ and $\mathbf{1}$ heavy right-handed neutrino $\nu^{}_{N}$ are added to the standard $3$ active neutrinos framework
\footnote{The reason why we consider the ($3$+$\mathbbm{1}$+$\mathbf{1}$) scenario is that current cosmological observations favored the existence of at most one species of light sterile neutrino, and for simplicity, we also introduce only one species of heavy sterile neutrino to illustrate the indirect unitarity violation effects. However, it is worth to mention that just one heavy right-handed neutrino is not enough to generate the neutrino masses. To accommodate the neutrino masses with the seesaw mechanism, one need to further introduce the Higgs triplet \cite{Minimal} or another generation(s) of heavy right-handed neutrino(s). A more general ($3$+$\mathbbm{1}$+$\mathbf{N}$) scenario is briefly discussed in the appendix.}.  In the ($3$+$\mathbbm{1}$+$\mathbf{1}$) scenario, the full picture of the neutrino mixing should be described by a $5 \times 5$ unitary matrix $V$
\begin{equation}
\left ( \begin{matrix} \nu^{}_{e} \cr \nu^{}_{\mu} \cr \nu^{}_{\tau} \cr \nu^{}_{s} \cr \nu^{}_{N} \cr \end{matrix} \right ) \; = \; \left ( \begin{matrix} V^{}_{e1} & V^{}_{e2} & V^{}_{e3} & V^{}_{e4} & V^{}_{e5} \cr V^{}_{\mu 1} & V^{}_{\mu 2} & V^{}_{\mu 3} & V^{}_{\mu 4} & V^{}_{\mu 5} \cr V^{}_{\tau 1} & V^{}_{\tau 2} & V^{}_{\tau 3} & V^{}_{\tau 4} & V^{}_{\tau 5} \cr V^{}_{s1} & V^{}_{s2} & V^{}_{s3} & V^{}_{s4} & V^{}_{s5} \cr V^{}_{N1} & V^{}_{N2} & V^{}_{N3} & V^{}_{N4} & V^{}_{N5} \cr \end{matrix} \right ) \; \left ( \begin{matrix} \nu^{}_{1} \cr \nu^{}_{2} \cr \nu^{}_{3} \cr \nu^{}_{4} \cr \nu^{}_{5} \cr \end{matrix} \right ) \; ,
%      (1)
\end{equation}
where $\nu^{}_{4}$ and $\nu^{}_{5}$ are corresponding mass eigenstates of the light and the heavy sterile neutrinos. Here we restrict us to the typical neutrino oscillation process $\nu^{}_{\alpha} \rightarrow \nu^{}_{\beta}$ where both the production of $\nu^{}_{\alpha}$ and the detection of $\nu^{}_{\beta}$ are via the charged-current interaction. Then the neutrino oscillation probability in vacuum can be written as \cite{Antusch}
\begin{eqnarray}
P \; ( \stackrel{(-)}{\nu}^{}_{\alpha} \rightarrow  \stackrel{(-)}{\nu}^{}_{\beta} ) & = & \frac{1}{\left ( \sum^{}_{i=1,2,3,4} | V^{}_{\alpha i} |^2 \right ) \left ( \sum^{}_{i=1,2,3,4} | V^{}_{\beta i} |^2 \right )} \left  \{ \left | \sum^{}_{i=1,2,3,4} V^{*}_{\alpha i} V^{}_{\beta i} \right |^2 \right. \nonumber\\
& & \left.  - 4 \sum^{}_{j > i} {\rm Re} \left [ V^{}_{\alpha i} V^{}_{\beta j} V^{*}_{\alpha j} V^{*}_{\beta i} \right ] \sin^2 \Delta^{}_{ji} \pm 2 \sum^{}_{j > i} {\rm Im} \left [ V^{}_{\alpha i} V^{}_{\beta j} V^{*}_{\alpha j} V^{*}_{\beta i} \right ] \sin2\Delta^{}_{ji} \right \} \nonumber\\
& = & \frac{1}{( 1 - | V^{}_{\alpha 5} |^2 ) ( 1 -| V^{}_{\beta 5} |^2 )} \left \{ \left | \delta^{}_{\alpha \beta} - V^{*}_{\alpha 5} V^{}_{\beta 5} \right |^2 \right. \nonumber\\
& & \left. - 4 {\rm Re} \left [ V^{}_{\alpha 1} V^{}_{\beta 2} V^{*}_{\alpha 2} V^{*}_{\beta 1} \right ] \sin^2 \Delta^{}_{21} \pm 2 {\rm Im} \left [ V^{}_{\alpha 1} V^{}_{\beta 2} V^{*}_{\alpha 2} V^{*}_{\beta 1} \right ] \sin2\Delta^{}_{21} \right. \nonumber\\
& & \left. - 4 {\rm Re} \left [ V^{}_{\alpha 1} V^{}_{\beta 3} V^{*}_{\alpha 3} V^{*}_{\beta 1} \right ] \sin^2 \Delta^{}_{31} \pm 2 {\rm Im} \left [ V^{}_{\alpha 1} V^{}_{\beta 3} V^{*}_{\alpha 3} V^{*}_{\beta 1} \right ] \sin2\Delta^{}_{31} \right. \nonumber\\
& & \left. - 4 {\rm Re} \left [ V^{}_{\alpha 2} V^{}_{\beta 3} V^{*}_{\alpha 3} V^{*}_{\beta 2} \right ] \sin^2 \Delta^{}_{32} \pm 2 {\rm Im} \left [ V^{}_{\alpha 2} V^{}_{\beta 3} V^{*}_{\alpha 3} V^{*}_{\beta 2} \right ] \sin2\Delta^{}_{32} \right. \nonumber\\
& & \left. - 4 {\rm Re} \left [ V^{}_{\alpha 1} V^{}_{\beta 4} V^{*}_{\alpha 4} V^{*}_{\beta 1} \right ] \sin^2 \Delta^{}_{41} \pm 2 {\rm Im} \left [ V^{}_{\alpha 1} V^{}_{\beta 4} V^{*}_{\alpha 4} V^{*}_{\beta 1} \right ] \sin2\Delta^{}_{41} \right. \nonumber\\
& & \left. - 4 {\rm Re} \left [ V^{}_{\alpha 2} V^{}_{\beta 4} V^{*}_{\alpha 4} V^{*}_{\beta 2} \right ] \sin^2 \Delta^{}_{42} \pm 2 {\rm Im} \left [ V^{}_{\alpha 2} V^{}_{\beta 4} V^{*}_{\alpha 4} V^{*}_{\beta 2} \right ] \sin2\Delta^{}_{42} \right. \nonumber\\
& & \left. - 4 {\rm Re} \left [ V^{}_{\alpha 3} V^{}_{\beta 4} V^{*}_{\alpha 4} V^{*}_{\beta 3} \right ] \sin^2 \Delta^{}_{43} \pm 2 {\rm Im} \left [ V^{}_{\alpha 3} V^{}_{\beta 4} V^{*}_{\alpha 4} V^{*}_{\beta 3} \right ] \sin2\Delta^{}_{43} \right \} \; ,
%      (2)
\end{eqnarray}
which in general consists of six CP-conserving oscillatory terms and six CP-violating oscillatory terms. Here $\Delta^{}_{ji} \simeq 1.27 \Delta m^2_{ji} L / E$ with $\Delta m^2_{ji} \equiv m^{2}_{j} - m^{2}_{i}$ is the neutrino mass-squared difference in ${\rm eV}^2$, $L$ is the baseline from the source to the detector in meters and $E$ is the neutrino or antineutrino energy in MeV. The Greek letters $\alpha$, $\beta$ are the flavor indices e, $\mu$ and $\tau$, while the Latin letters $i$, $j$ are the mass indices. Note that the indices $i$, $j$ in Eq. (2) run over only the {\bf light} neutrinos (both the active and the sterile) which can be kinematically produced in neutrino oscillation experiments and the normalization factor $1 /( \sum^{}_{i=1,2,3,4} | V^{}_{\alpha i} |^2 ) ( \sum^{}_{i=1,2,3,4} | V^{}_{\beta i} |^2 )$ ensures that the total rate $P(W^{+}_{} \rightarrow \bar{l}_{\alpha}^{} \nu_{\alpha}^{}) \equiv \sum_{i}^{} |A(W^{+}_{} \rightarrow \bar{l}_{\alpha}^{} \nu_{i}^{})|^2 = 1$ (at the source) and $P(\nu_{\beta}^{} W^{-}_{} \rightarrow l_{\beta}^{}) \equiv \sum_{i}^{} |A(\nu_{i}^{} W^{-}_{} \rightarrow l_{\beta}^{})|^2 = 1$ (at the detector).

The possible effects of both the direct and the indirect unitarity violation in neutrino oscillation experiments have been discussed in many previous papers. For example, in the presence of heavy sterile neutrinos, the oscillation probabilities have the property $P \; ( \stackrel{(-)}{\nu}^{}_{\alpha} \rightarrow  \stackrel{(-)}{\nu}^{}_{\beta} ) \neq \delta^{}_{\alpha \beta}$ in the limit $L \rightarrow 0$ which is well known as the ``zero-distance'' effect \cite{zero}. We can clearly see from Eq. (2) that such effect will not take place if there exist only light sterile neutrinos. Therefore it would be a definite signal of the indirect unitarity violation if the ``zero-distance'' effect can be observed in future neutrino oscillation experiments. To obtain the best sensitivities to certain parameters (mixing angles or CP-violating phases) of the direct or the indirect unitarity violation, plenty of works have been done to find the optimum setups by choosing appropriate neutrino source, oscillation channels and baselines or by proceeding a combined analysis of the data from different baselines where the matter effect may play quite different roles \cite{UV}.

However, in this paper, we focus on the reactor antineutrino oscillation experiments where only CP-conversing terms are involved in the electron antineutrino survival probability 
\begin{eqnarray}
P(\bar{\nu}^{}_{e} \rightarrow \bar{\nu}^{}_{e}) & = & \frac{1}{( 1 - | V^{}_{e 5} |^2 )^2} \left \{ \left ( 1 - |V^{}_{e5}|^2 \right )^2 \right. \nonumber\\
& & \left. - 4 |V^{}_{e1}|^2 |V^{}_{e2}|^2 \sin^2 \Delta^{}_{21} - 4 |V^{}_{e1}|^2 |V^{}_{e3}|^2 \sin^2 \Delta^{}_{31} - 4 |V^{}_{e2}|^2 |V^{}_{e3}|^2 \sin^2 \Delta^{}_{32} \right. \nonumber\\
& & \left. - 4 |V^{}_{e1}|^2 |V^{}_{e4}|^2 \sin^2 \Delta^{}_{41} - 4 |V^{}_{e2}|^2 |V^{}_{e4}|^2 \sin^2 \Delta^{}_{42} - 4 |V^{}_{e3}|^2 |V^{}_{e4}|^2 \sin^2 \Delta^{}_{43} \right \} \; .
%      (3)
\end{eqnarray}
The standard formula for only three active neutrinos can be easily reproduced by simply choosing $|V^{}_{e4}| = |V^{}_{e5}| = 0$ in Eq. (3). For the ($3$+$\mathbbm{1}$) or ($3$+$\mathbf{1}$) scenario where only one light or one heavy sterile neutrino is added, the corresponding survival probabilities can be obtained by taking $|V^{}_{e5}| = 0$ or $|V^{}_{e4}| = 0$ respectively.

An elegant parametrization has been proposed to parametrize the full unitary mixing matrix \cite{Xing2012}. In the ($3$+$\mathbbm{1}$+$\mathbf{1}$) scenario, the $5 \times 5$ matrix $V$ in Eq. (1) can be similarly decomposed as
\begin{equation}
V \; = \; \left ( \begin{matrix} {\bm 1} & {\bm 0} \cr {\bm 0} & U^{}_{0} \cr \end{matrix} \right ) \left ( \begin{matrix} A & R \cr S & B \cr \end{matrix} \right ) \left ( \begin{matrix} V^{}_{0} & {\bm 0} \cr {\bm 0} & {\bm 1} \cr \end{matrix} \right ) \; ,
%      (4)
\end{equation}
in which $V^{}_{0}$ and $A$ are $3 \times 3$ matrices, $U^{}_{0}$ and $B$ are $2 \times 2$ matrices, $R$ is a $3 \times 2$ matrix, $S$ is a $2 \times 3$ matrix while $\bm{0}$ and $\bm{1}$ stand respectively for the zero and identity matrices. These matrices are parametrized as
\begin{eqnarray}
\left ( \begin{matrix} {\bm 1} & {\bm 0} \cr {\bm 0} & U^{}_{0} \cr \end{matrix} \right ) & = & {\cal O}^{}_{45} \; , \nonumber\\
\left ( \begin{matrix} A & R \cr S & B \cr \end{matrix} \right ) & = & {\cal O}^{}_{35} {\cal O}^{}_{25} {\cal O}^{}_{15} {\cal O}^{}_{34} {\cal O}^{}_{24} {\cal O}^{}_{14} \; , \nonumber\\
\left ( \begin{matrix} V^{}_{0} & {\bm 0} \cr {\bm 0} & {\bm 1} \cr \end{matrix} \right ) & = & {\cal O}^{}_{23} {\cal O}^{}_{13} {\cal O}^{}_{12} \; ,
%      (5)
\end{eqnarray}
where ten ${\cal O}^{}_{ij}$ (for $1\leq i < j \leq 5$) are two-dimensional rotation matrices in the five-dimensional complex space whose explicit expressions can be found in Ref. \cite{Xing2012}. One can easily see from this parametrization that the matrices $V^{}_{0}$ and $U^{}_{0}$ are unitary while $A$, $B$, $R$, $S$ are not. The production $A V^{}_{0}$ can be regarded as the effective $3 \times 3$ MNSP matrix in this ($3$+$\mathbbm{1}$+$\mathbf{1}$) scenario which is in general non-unitary.

An apparent advantage of this parametrization is that all the five moduli $|V^{}_{ei}|$ that are involved in Eq. (3) have very concise expressions:
\begin{eqnarray}
|V^{}_{e1}| & = & c^{}_{12} c^{}_{13} c^{}_{14} c^{}_{15} \; , \nonumber\\
|V^{}_{e2}| & = & s^{}_{12} c^{}_{13} c^{}_{14} c^{}_{15} \; , \nonumber\\
|V^{}_{e3}| & = & s^{}_{13} c^{}_{14} c^{}_{15} \; , \nonumber\\
|V^{}_{e4}| & = & s^{}_{14} c^{}_{15} \; , \nonumber\\
|V^{}_{e5}| & = & s^{}_{15} \; ,
%      (6)
\end{eqnarray}
where $c^{}_{ij} \equiv \cos\theta^{}_{ij}$ and $s^{}_{ij} \equiv \sin\theta^{}_{ij}$ with $ij = 12, 13, 14$ and $15$. Here $\theta^{}_{14}$  stands for the mixing between the light sterile neutrino and the active neutrinos while $\theta^{}_{15}$ stands for the mixing between the heavy sterile neutrino and the active neutrinos. We can clearly find in Eq. (6) that if $\theta^{}_{14} = 0$ then we have $|V^{}_{e4}| = 0$, while $|V^{}_{e5}| = 0$ can be easily obtained by taking $\theta^{}_{15} = 0$. The survival probability  $P(\bar{\nu}^{}_{e} \rightarrow \bar{\nu}^{}_{e})$ is then given by
\begin{eqnarray}
P(\bar{\nu}^{}_{e} \rightarrow \bar{\nu}^{}_{e}) & = & 1 - P^{}_{21} - P^{}_{31} - P^{}_{32} - P^{}_{41} - P^{}_{42} - P^{}_{43} \nonumber\\
& = & 1 - 4 s^{2}_{12} c^{2}_{12} c^{4}_{13} c^{4}_{14} \sin^2 \Delta^{}_{21} - 4 c^{2}_{12} s^{2}_{13} c^{2}_{13} c^{4}_{14}  \sin^2 \Delta^{}_{31} - 4 s^{2}_{12} s^{2}_{13} c^{2}_{13} c^{4}_{14}  \sin^2 \Delta^{}_{32} \nonumber\\
& & - 4 c^{2}_{12} c^{2}_{13} s^{2}_{14} c^{2}_{14} \sin^2 \Delta^{}_{41} - 4 s^{2}_{12} c^{2}_{13} s^{2}_{14} c^{2}_{14} \sin^2 \Delta^{}_{42} - 4 s^{2}_{13} s^{2}_{14} c^{2}_{14} \sin^2 \Delta^{}_{43} \; .
%      (7)
\end{eqnarray}
One may immediately find from Eq. (7) that the mixing angle $\theta^{}_{15}$ is not shown in the electron antineutrino survival probability which implies that the reactor experiment is almost insensitive to the indirect unitarity violation induced by the heavy sterile neutrinos. An exception is that the indirect unitarity violation of the MNSP matrix will result in corrections to the cross sections of both the charged-current and the neutral-current interactions \cite{Antusch}. However, precise calculation of the reactor antineutrino spectrum, exact value of the detector efficiency and accurate absolute energy scale calibration in the detectors are required for probing this minor effect. In this paper we will only discuss the antineutrino survival probability itself and focus on the direct unitarity violation effects in the reactor experiments induced by sub-eV sterile neutrinos. Therefore the following discussions are simply carried out in the ($3$+$\mathbbm{1}$) scenario. 

Before ending this section, it is worth to mention that in the ($3$+$\mathbbm{1}$+$\mathbf{1}$) scenario, altogether 14 new independent mass and mixing parameters are introduced 
\footnote{These 14 additional parameters consist of 7 mixing angles (of which $\theta^{}_{14}$, $\theta^{}_{24}$ and $\theta^{}_{34}$ describe the mixing between three active neutrinos and the light sterile neutrino, $\theta^{}_{15}$, $\theta^{}_{25}$ and $\theta^{}_{35}$ describe the mixing between three active neutrinos and the heavy sterile neutrino, and $\theta^{}_{45}$ describes the mixing between the light and the heavy sterile neutrinos as one can clearly see in Eq. (5)), 5 phases and 2 sterile neutrino masses.}
, but only two of them ($\theta^{}_{14}$ and $\Delta m^{2}_{41}$) are relevant to the electron antineutrino survival probabilities. And we will show in the next section that the reactor experiments can provide definite signals for each of them. 

\section{Search for the Sub-\lowercase{e}V Sterile Neutrinos}

Now we focus on the reactor antineutrino oscillation experiment with three different baselines: $L^{}_{1} = 500$ m, $L^{}_{2} = 2$ km and $L^{}_{3} = 60$ km. The six oscillatory terms in Eq. (7) may behave very different at the three different baselines, which provide the opportunity to distinguish the unitarity violation parameters from the standard ones. The combination of the DayaBay experiment \cite{DayaBay} and the upcoming JUNO (Jiangmen Underground Neutrino Observatory, formerly known as Daya Bay II) experiment \cite{JUNO} is just of this type so is the RENO experiment \cite{RENO} combined with the proposed RENO-50 reactor experiment \cite{RENO-50}. Studying from two aspects: the rate analysis and the spectral analysis, we are going to discuss the sensitivities of this kind of reactor experiment to the parameters $\theta^{}_{14}$ and $\Delta m^{2}_{41}$ in detail. In the following, $\theta^{}_{12} = 33.65^\circ$, $\theta^{}_{13} = 8.9^\circ$, $\Delta m^{2}_{21} = 7.6 \times 10^{-5}_{} ~ {\rm eV}^2$, and $|\Delta m^{2}_{31}| = 2.4 \times 10^{-3}_{} ~ {\rm eV}^2$ are chosen as default unless otherwise specified. 

\begin{figure}
\begin{center}
\vspace{0cm}
\includegraphics[scale=0.6, angle=0, clip=0]{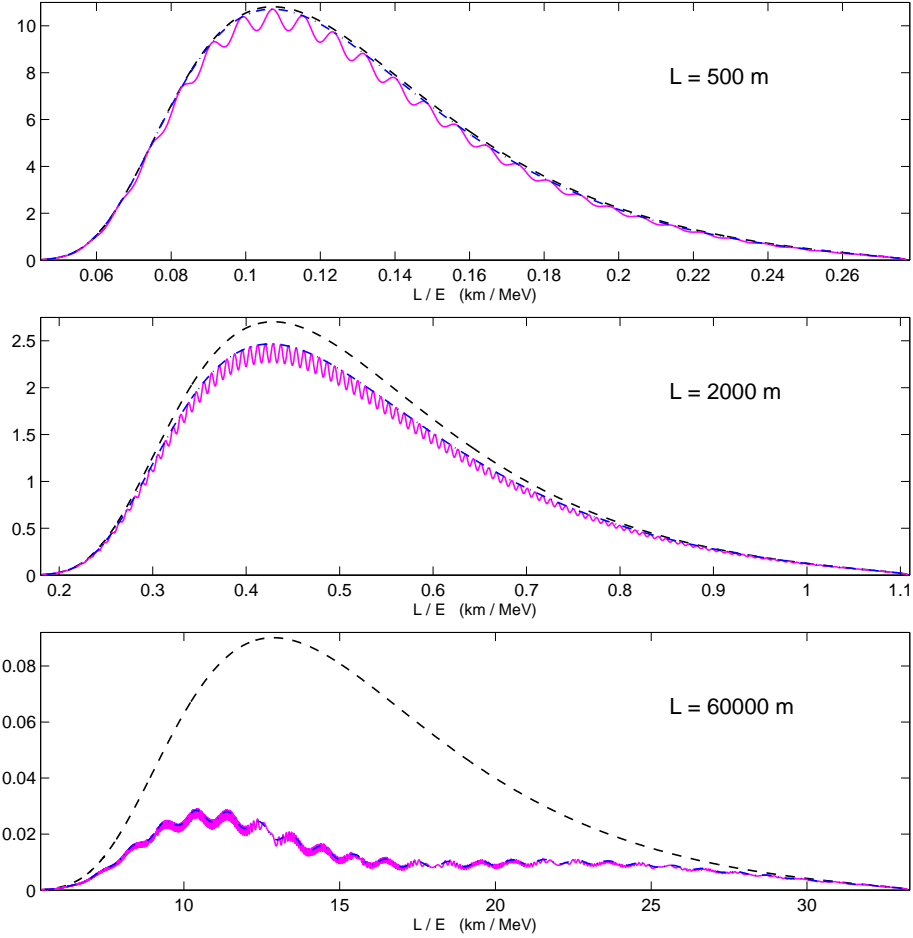}
\vspace{-0.3cm}
\caption{Reactor antineutrino spectra in the $L / E$ space for no oscillation (dashed line), the standard three active neutrinos case (dash-dotted line) and the ($3$+$\mathbbm{1}$) scenario with $\Delta m^{2}_{41} = 0.3 ~ {\rm eV}^2$ and $|V^{}_{e4}|^2 = 0.01$ (solid line) at the baselines of 500 m, 2 km and 60 km respectively.}
\end{center}
\end{figure}

\begin{figure}
\begin{center}
\vspace{0cm}
\includegraphics[scale=0.6, angle=0, clip=0]{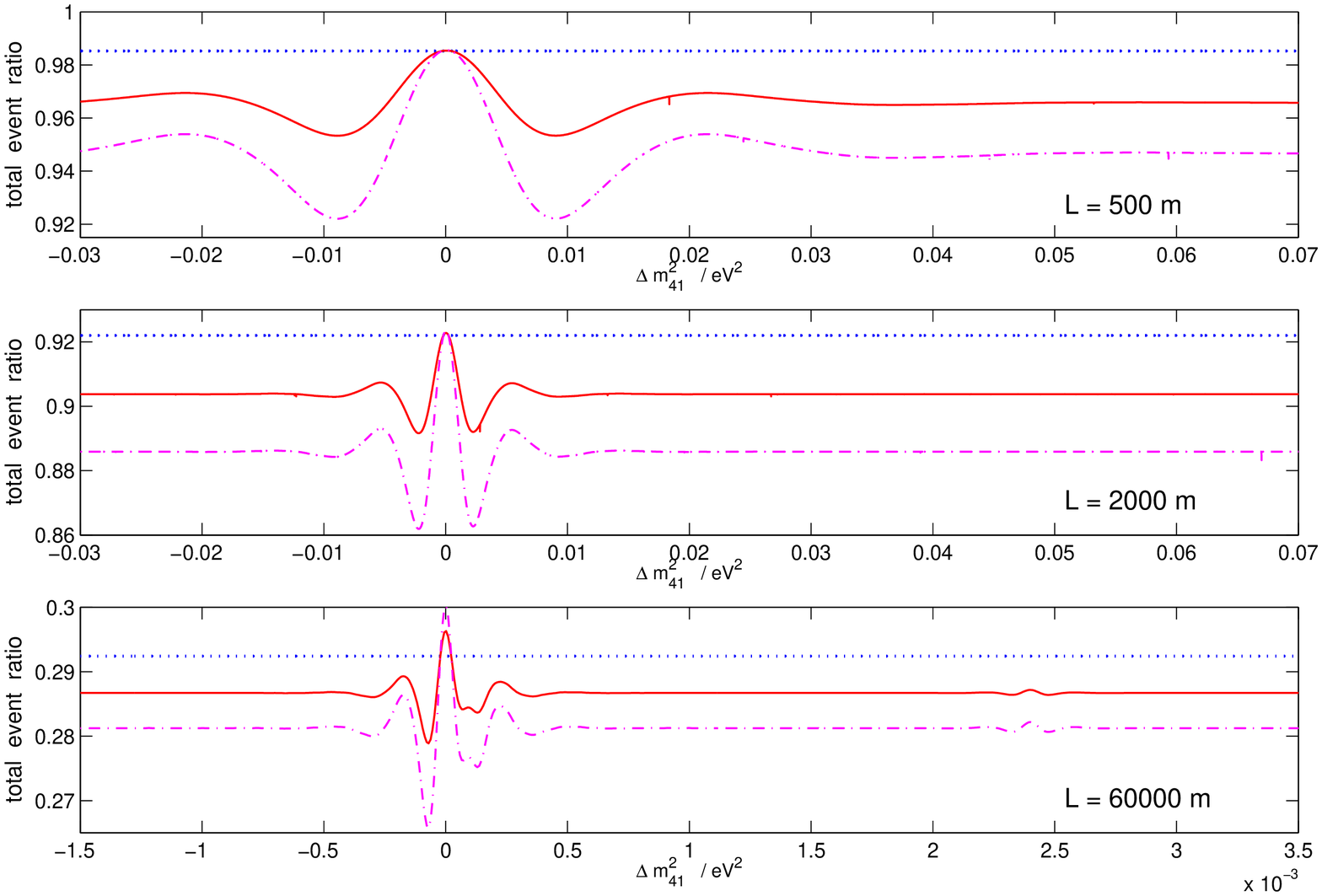}
\vspace{0cm}
\caption{Ratio of the total observed events to the no oscillation expectation as a function of $\Delta m^{2}_{41}$ at the baselines of 500 m, 2 km and 60 km respectively. The solid line stands for the (3+$\mathbbm{1}$) scenario with $|V^{}_{e4}|^2 = 0.01$, the dash-dotted line for the (3+$\mathbbm{1}$) scenario with $|V^{}_{e4}|^2 = 0.02$ and the dotted line for the standard three active neutrinos case.}
\end{center}
\end{figure}

\subsection{Rate Analysis}

For a reactor neutrino experiment, the observed electron antineutrino spectrum $F$ at a baseline $L$, in the $L / E$ space can be written as \cite{Zhan2008}
\begin{eqnarray}
F (L / E) & = & \phi(E) \sigma(E) P(\bar{\nu}^{}_{e} \rightarrow \bar{\nu}^{}_{e}) \frac{E^2}{L} \; ,
%      (8)
\end{eqnarray}
where $E$ is the electron antineutrino ($\bar{\nu}^{}_{e}$) energy, $\sigma(E)$ is the interaction cross section of $\bar{\nu}^{}_{e}$ with matter \cite{sigma} and $\phi(E)$ is the flux of $\bar{\nu}^{}_{e}$ from the reactor \cite{flux}.
Taking the baseline $L$ to be 500 m, 2 km and 60 km respectively, the observed neutrino spectra in the $L / E$ space are shown in Fig. 1 where the solid line stands for the (3+$\mathbbm{1}$) scenario with $\Delta m^{}_{41} = 0.3 ~ {\rm eV}^2$ and $|V^{}_{e4}|^2 = 0.01$, the dash-dotted line for the standard three active neutrinos case and the dashed line is the no oscillation spectrum for comparison. With current energy resolution, the oscillatory frequencies of $P^{}_{41}$, $P^{}_{42}$ and $P^{}_{43}$ at the baseline $L^{}_{3} = 60$ km are rather high, thus their oscillatory behaviors are highly suppressed and only the averaged spectrum can be detected. 

The total number of events observed in the detector can be calculated by integrating the antineutrino flux over the energy. Fig. 2 shows the total event ratio which is the ratio of the total energy-integrated events to the no oscillation expectation as a function of $\Delta m^{2}_{41}$ at the three different baselines. In this figure, the solid line stands for the (3+$\mathbbm{1}$) scenario with $|V^{}_{e4}|^2 = 0.01$, the dash-dotted line for the (3+$\mathbbm{1}$) scenario with $|V^{}_{e4}|^2 = 0.02$ and the dotted line for the standard three active neutrinos case. One can see that the total event ratio is sensitive only to very small $\Delta m^{2}_{41}$. The reason is that if $\Delta^{}_{ji}$ is large, $\sin^2 \Delta^{}_{ji}$ oscillates very fast with the varying of $E$, and therefore is fully averaged when integrated over the energy. We can see from Fig. 2, if $\Delta m^{2}_{41} > 0.05 ~ {\rm eV}^2$,  the total event ratio is almost independent of $\Delta m^{2}_{41}$ at all the three baselines while still sensitive to the sterile-active mixing angle $\theta^{}_{14}$.

Compared with the standard three active neutrinos case, the existence of additional light sterile neutrinos will in generally lead to additional depression of the total event ratio and can mimic the signal of $\theta^{}_{13}$ if it is extracted from the rate analysis at a single baseline \cite{reactor sterile}. To see this point more clearly, Fig. 3 shows the contour lines of the the total event ratio in the $\theta^{}_{13}$-$\theta^{}_{14}$ plane at the three different baselines. Instead of a definite value of $\theta^{}_{13}$, the measured total event ratio at any single baseline gives only possible ranges of $\theta^{}_{13}$ and $\theta^{}_{14}$ together with the correlation between these two mixing angles.

However, this situation can be basically changed for the multiple baselines reactor experiment, where there are usually detectors at the near site playing the role of calibrator and $\theta^{}_{13}$ is determined by comparing the event rates at the near and the far baselines. Fig. 4 shows the contour lines of the relative total event ratio in the $\theta^{}_{13}$-$\theta^{}_{14}$ plane at the baselines of 2 km and 60 km respectively, where the total event ratios at these two baselines are normalized by that at the baseline of $L^{}_{1} = 500$ m. We can find that the true value of $\theta^{}_{13}$ can be determined independently of $\theta^{}_{14}$ by the relative event rate at $L^{}_{2} = 2$ km. On the other hand, the relative event rate at $L^{}_{3} = 60$ km is jointly determined by the values of $\theta^{}_{13}$ and $\theta^{}_{14}$. It implies that if the total event ratio or the relative event rate at the third baseline around 60 km can be precisely measured in the upcoming JUNO or RENO-50 experiments, together with the well determined $\theta^{}_{13}$, we are able to draw information on the active-sterile mixing angle $\theta^{}_{14}$. It is worth to mention that, although we have typically set $\Delta m^{2}_{41} = 0.3 ~ {\rm eV}^2$ in plotting Fig. 4, the conclusion keeps unchanged for any $\Delta m^{2}_{41} > 0.05 ~ {\rm eV}^2$.

\begin{figure}
\begin{center}
\vspace{0cm}
\includegraphics[scale=0.49, angle=0, clip=0]{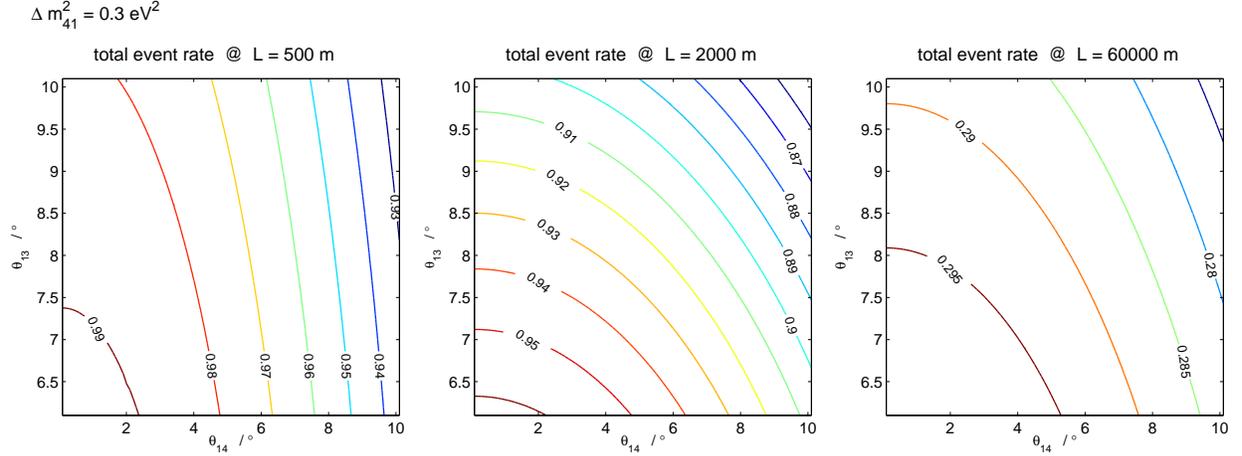}
\vspace{-0.3cm}
\caption{Contour lines of the total event ratio in the $\theta^{}_{13}$-$\theta^{}_{14}$ plane at the baselines of 500 m, 2 km and 60 km respectively. Although we have typically set $\Delta m^{2}_{41} = 0.3 ~ {\rm eV}^2$ in plotting these contour figures, the results are almost exactly the same for any $\Delta m^{2}_{41} > 0.05 {\rm eV}^2$.}
\end{center}
\end{figure}

\begin{figure}
\begin{center}
\vspace{0cm}
\includegraphics[scale=0.6, angle=0, clip=0]{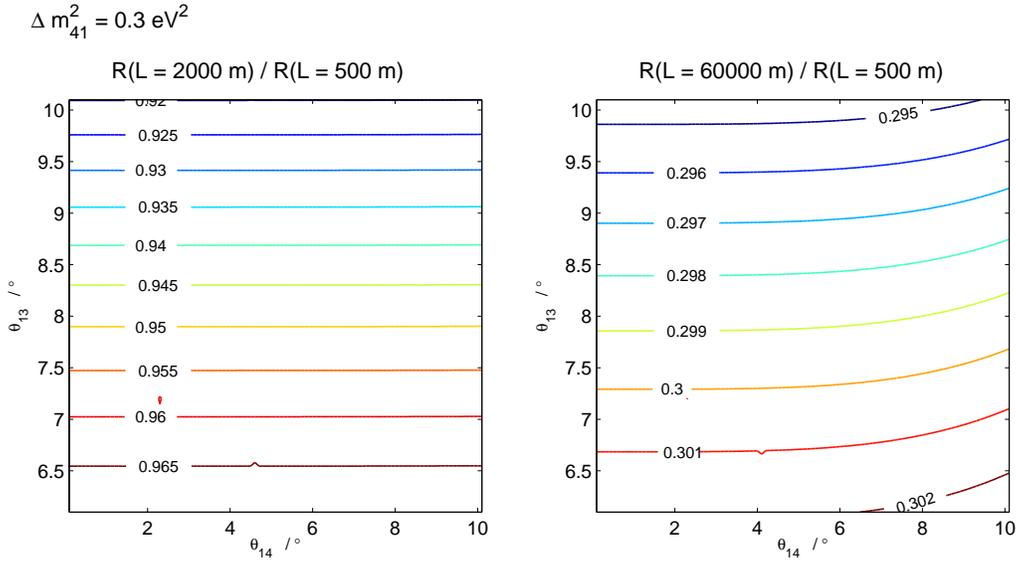}
\vspace{0cm}
\caption{Contour lines of the relative total event ratio in the $\theta^{}_{13}$-$\theta^{}_{14}$ plane at the baselines of 2 km and 60 km respectively. Although we have typically set $\Delta m^{2}_{41} = 0.3 ~ {\rm eV}^2$ in plotting these contour figures, the results are almost exactly the same for any $\Delta m^{2}_{41} > 0.05 {\rm eV}^2$.}
\end{center}
\end{figure}

Above conclusions are theoretically understandable as the result of the different behaviors of the six oscillatory terms in Eq. (7) at different baselines. Suppose $\Delta m^{2}_{41} > 0.05 {\rm eV}^2$ is always satisfied, the three oscillatory terms $\sin^2 \Delta^{}_{41}$, $\sin^2 \Delta^{}_{42}$ and $\sin^2 \Delta^{}_{43}$ are fully averaged ($\approx 1/2$) at all the three baselines. At the near site $L^{}_{1} = 500$ m, the energy-averaged antineutrino survival probability can be  approximately written as
\begin{eqnarray}
P(L = 500 ~ {\rm m}) & \approx & 1 - 2 s^{2}_{14} c^{2}_{14} = c^{4}_{14} + s^{4}_{14} \; ,
%      (9)
\end{eqnarray}
where the three terms $P^{}_{21}$, $P^{}_{31}$ and $P^{}_{32}$ are neglected because of the smallness of $\sin^2 \Delta^{}_{21}$, $s^{2}_{13} \sin^2 \Delta^{}_{31}$ and $s^{2}_{13} \sin^2 \Delta^{}_{32}$ at this baseline. At the baseline $L^{}_{2} = 2$ km, the survival probability can be  approximately written as
\begin{eqnarray}
P(L = 2000 ~ {\rm m}) & \approx & 1 - 2 s^{2}_{14} c^{2}_{14} - c^{4}_{14} \sin^2 2 \theta^{}_{13} \left [ \sin^2 \Delta^{}_{31} \right ]_{2000 {\rm m}}  \nonumber\\
& = & c^{4}_{14} \left ( 1 - \sin^2 2 \theta^{}_{13} \left [ \sin^2 \Delta^{}_{31} \right ]_{2000 {\rm m}} \right ) + s^{4}_{14} \; ,
%      (10)
\end{eqnarray}
where $\left [ \sin^2 \Delta^{}_{31} \right ]_{2000 {\rm m}}$ stands for the energy-averaged value of $\sin^2 \Delta^{}_{31}$ at $L^{}_{2} = 2$ km and terms proportional to $\sin^2 \Delta^{}_{21}$ are safely neglected. Then the relative event rate at $L^{}_{2} = 2$ km can be estimated by 
\begin{eqnarray}
\frac{P(L = 2000 ~ {\rm m})}{P(L = 500 ~ {\rm m})} & \approx & 1 - \left ( 1 - s^{4}_{14} \right ) \sin^2 2 \theta^{}_{13} \left [ \sin^2 \Delta^{}_{31} \right ]_{2000 {\rm m}} + {\cal O} (s^{8}_{14}) \; .
%      (11)
\end{eqnarray}
The leading terms that are dependent of $\theta^{}_{14}$ in Eq. (11) are proportional to $s^{4}_{14}$ and are further suppressed by the small factor $\sin^2 2 \theta^{}_{13}$. This clearly explained that the estimate of $\theta^{}_{13}$ by the combined rate analysis at the two baselines $L^{}_{1} = 500$ m and $L^{}_{2} = 2$ km is nearly independent of the value of $\theta^{}_{14}$.

At the baseline of $L^{}_{3} = 60$ km, we can infer from the third plot of Fig. 2 that all the $\sin^2 \Delta^{}_{ji}$ terms with $\Delta m^{2}_{ji} > 10^{-3} ~ {\rm eV}^2$ are fully averaged out. Therefore $P^{}_{21}$ is the dominate oscillatory term at this baseline and the energy-averaged electron antineutrino survival probability should be approximately written as
\begin{eqnarray}
P(L = 60000 ~ {\rm m}) & \approx & 1 - 2 s^{2}_{13} c^{2}_{13} c^{4}_{14} - 2 s^{2}_{14} c^{2}_{14} - c^{4}_{13} c^{4}_{14} \sin^2 2 \theta^{}_{12} \left [ \sin^2 \Delta^{}_{21} \right ]_{60000 {\rm m}} \nonumber\\
& = & c^{4}_{13} c^{4}_{14} \left ( 1 - \sin^2 2 \theta^{}_{12} \left [ \sin^2 \Delta^{}_{21} \right ]_{60000 {\rm m}} \right ) + s^{4}_{13} c^{4}_{14} + s^{4}_{14} \; ,
%      (12)
\end{eqnarray}
where $\left [ \sin^2 \Delta^{}_{21} \right ]_{60000 {\rm m}}$ is the energy-averaged value of $\sin^2 \Delta^{}_{21}$ at $L^{}_{3} = 60$ km. Then we have
\begin{eqnarray}
\frac{P(L = 60000 ~ {\rm m})}{P(L = 500 ~ {\rm m})} & \approx & c^{4}_{13} \left [ 1 - \left ( 1 - s^{4}_{14} \right ) \sin^2 2 \theta^{}_{12} \left [ \sin^2 \Delta^{}_{21} \right ]_{60000 {\rm m}} \right ] \nonumber\\
& & + s^{4}_{13} + \frac{1}{2} \sin^2 2\theta^{}_{13} s^{4}_{14} + {\cal O} (s^{8}_{14}) \; .
%      (13)
\end{eqnarray}

Figure 5 shows the total event ratio at the baseline of 60 km and the relative total event rate $P(L = 60000 ~ {\rm m})/P(L = 500 ~ {\rm m})$ as the functions of $\theta^{}_{14}$
\footnote{The small peeks/dips in Figs. 4 and 5 are caused by the numerical errors, which arise from the integrations. This kind of numerical errors might be accidentally amplified when calculating the ratio of two integrated event rates.}.
In order to get some information on the active-sterile mixing angle $\theta^{}_{14}$ or to put an upper limit on it, the total event ratio should be precisely determined to the level of ${\cal O}(10^{-3})$, or the relative event rate should be precisely measured to the level of ${\cal O}(10^{-4})$ at the third baseline around 60 km in future precision reactor experiments, and the uncertainties from other mixing parameters ($\theta^{}_{12}$, $\theta^{}_{13}$ and $\Delta m^{2}_{21}$) should also be well reduced to have the same precision.

\begin{figure}
\begin{center}
\vspace{0cm}
\includegraphics[scale=0.54, angle=0, clip=0]{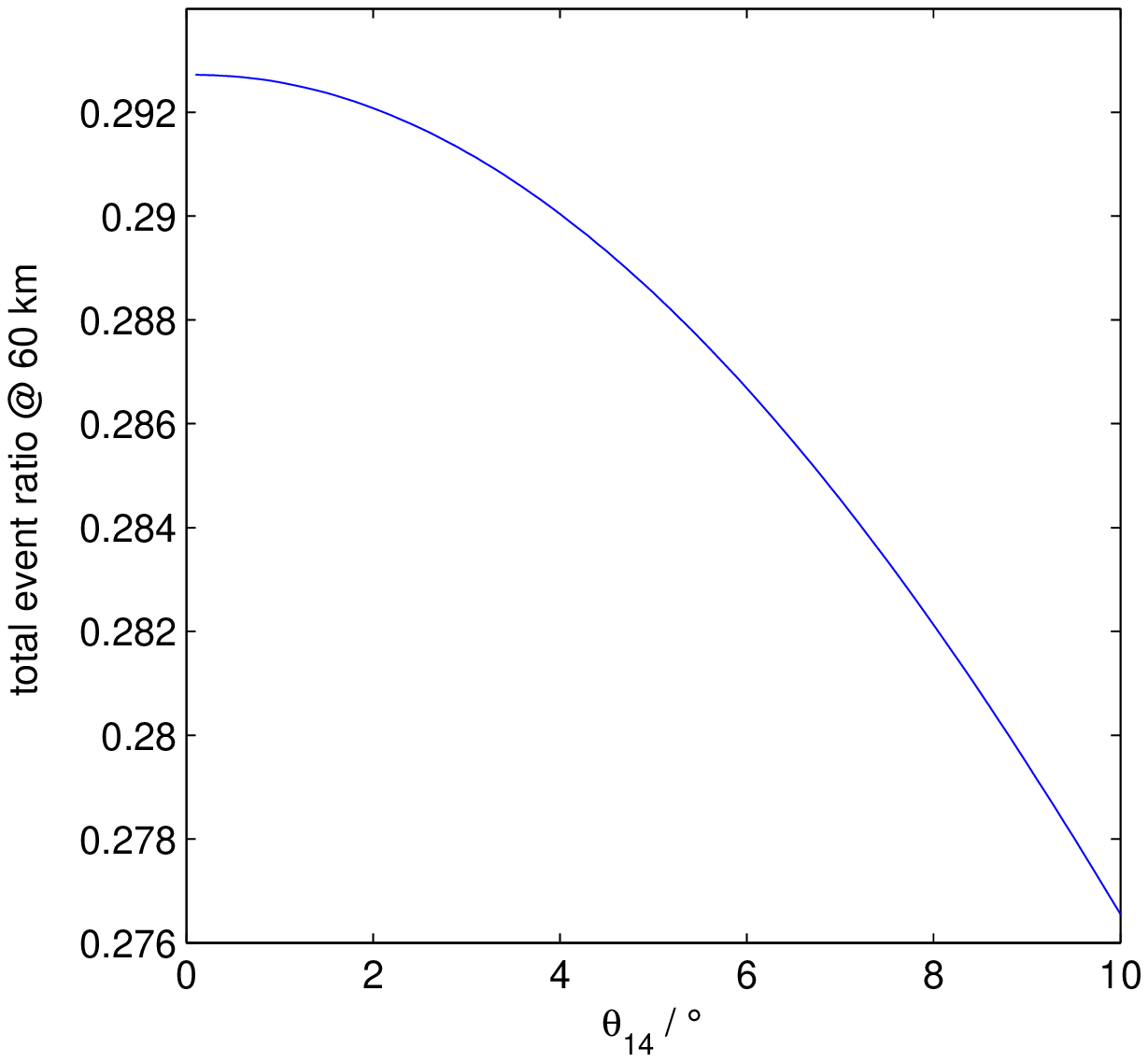}
\hspace{0.6cm}
\includegraphics[scale=0.54, angle=0, clip=0]{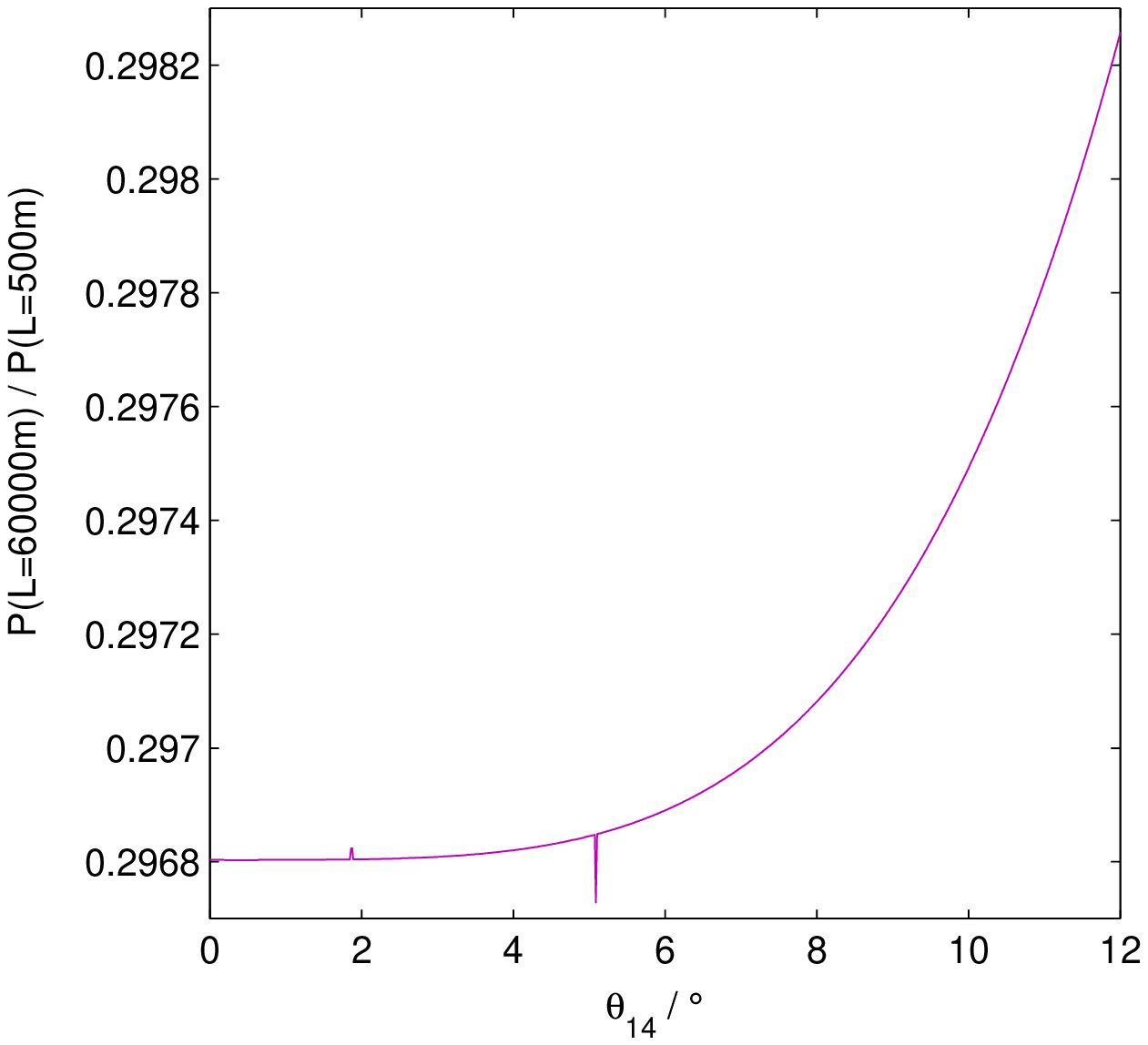}
\hspace{0.4cm}
\vspace{-0.1cm}
\caption{The total event ratio at the baseline of 60 km and the relative event rate $P(L = 60000 ~ {\rm m})/P(L = 500 ~ {\rm m})$ as the functions of $\theta^{}_{14}$, where $\theta^{}_{13} = 8.9^\circ$ has been chosen as an input.}
\end{center}
\end{figure}

\subsection{Spectral Analysis}

One can see from Eq. (7) that  three new oscillatory terms $P^{}_{41}$, $P^{}_{42}$ and $P^{}_{43}$ (i.e., three new oscillatory frequencies) are included in due to the existence of one light sterile neutrino $\nu^{}_{4}$. A direct measurement of the oscillatory behaviors of these three terms will certainly provide the direct evidence of the existence of such light sterile neutrinos. However the amplitudes of all these three oscillations are rather small (proportional to $s^{2}_{14}$). It has been found that comparing to a normal $L / E$ spectrum analysis, the Fourier analysis naturally separates the mass hierarchy information from uncertainties of the reactor antineutrino spectra and other mixing parameters, which is critical for very small oscillations. The frequency spectrum can be obtained by applying the following Fourier sine transformation (FST) and Fourier cosine transformation (FCT) to the $L / E$ spectra of the antineutrinos:
\begin{eqnarray}
FST(\omega) & = & \int^{t_{max}}_{t_{min}} F(t) \sin(\omega t ) {\rm d}t \; , \\
FCT(\omega) & = & \int^{t_{max}}_{t_{min}} F(t) \cos(\omega t ) {\rm d}t \; , 
%      (14)
\end{eqnarray}
where $\omega$ is the frequency. Here we set $\omega = \Delta m^{2}_{ji}$ just to be the mass-squared differences and $t = L / 2.54 E$ is the viable in $L / E$ space. In this convention, we can easily read the value of the corresponding $\Delta m^{2}_{ji}$ from the FST or FCT spectra.
We typically choose $\Delta m^{2}_{41} = 0.3 ~ {\rm eV}^{2}_{}$ and show in Fig. 6 the corresponding FST and FCT spectra at the three different baselines.

\begin{figure}
\begin{center}
\vspace{-1.8cm}
\includegraphics[scale=0.6, angle=0, clip=0]{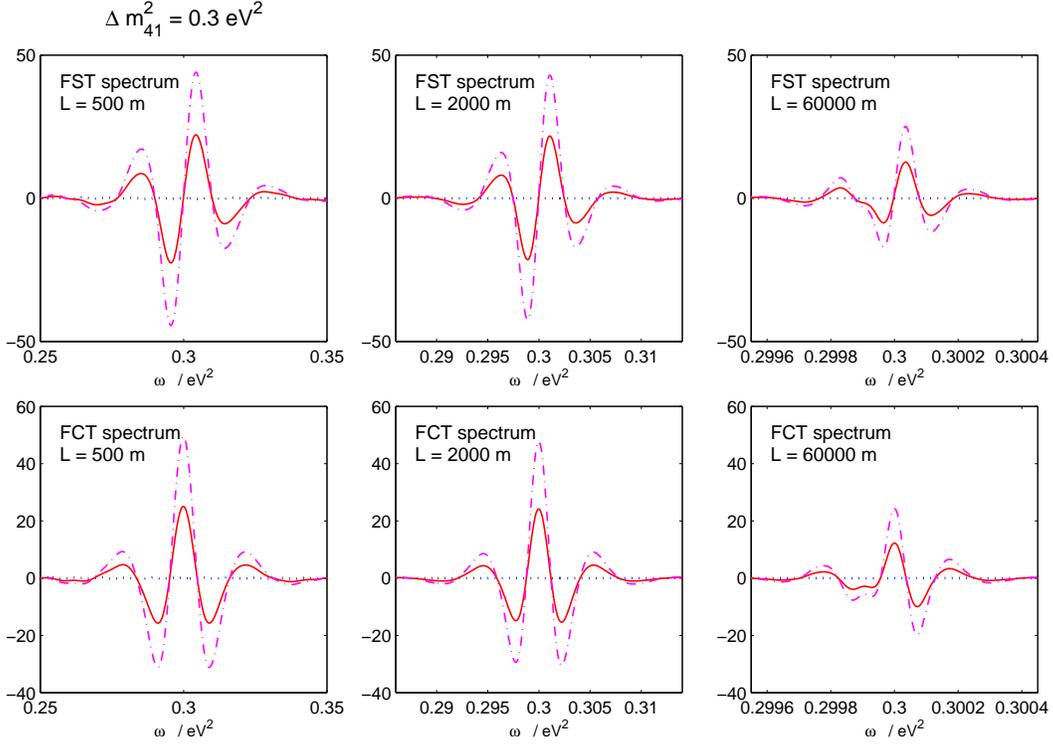}
\vspace{-0.3cm}
\caption{Fourier sine (FST) and cosine (FCT) transformation spectra at the baselines of 500 m, 2 km and 60 km respectively for the standard three active neutrinos case (dotted line) and (3+$\mathbbm{1}$) case with $|V^{}_{e4}|^2 = 0.01$ (solid line) or $|V^{}_{e4}|^2 = 0.02$ (dash-ditted line).}
\end{center}
\end{figure}

Whether the information of $\Delta m^{2}_{41}$ can be extracted from the spectra depend strongly on the energy resolution and the statistics. The simulation in Ref. \cite{resolution} suggests that the energy resolution $\delta E / E$ should be better than $0.68 \pi / \Delta^{}_{ji}$ so that the corresponding high frequency oscillatory behavior of $P^{}_{ji}$ is not completely suppressed. Taking $\Delta m^{2}_{41} \sim 0.3 ~ {\rm eV}^2$ and the antineutrino energy $E \sim 4$ MeV, we can than give a estimate of the required lowest energy resolutions: $4.49\%$ at the baseline $L^{}_{1} = 500$ m, $1.12 \%$ at $L^{}_{2} = 2$ km and $0.04 \%$ at $L^{}_{3} = 60$ km. Note that the larger $\Delta m^{2}_{41}$ we are aiming and the longer baseline we have chosen, the higher energy resolution are required. The upcoming JUNO experiment is aiming at a new high detector energy resolution of $3\% / \sqrt{E({\rm MeV})}$. If the near detectors at several hundred meters can be upgraded to the same high energy resolution, it is possible to find some clues of the sub-eV sterile neutrinos.

The absolute value of $\theta^{}_{14}$ is another crucial condition for this kind of measurement. One can infer from Eq. (7), the amplitude of the FST / FCT spectra of the three newly induced oscillatory terms $P^{}_{41}$, $P^{}_{42}$ and $P^{}_{43}$ is proportional to $c^{2}_{13} \sin^2 \theta^{}_{14}$. For smaller $\sin^2 \theta^{}_{14}$, the main peak become less significant. In order to clearly identify the FST / FCT spectra of $P^{}_{41}$, $P^{}_{42}$ and $P^{}_{43}$, the main peak is required to be at least twice higher than that of the noise which could be either the spectra of other oscillatory terms or the statistical fluctuations. Obviously, fewer number of events will induce larger statistical fluctuations, more noisy peaks and valleys in the FCT and FST spectra and hence reduce the discovery probability. Considering the fact that $\sin^2 \theta^{}_{14}$ is constrained to be at most a few percents, large statistics is need, which means massive detectors as well as powerful reactors are highly required.

Of course, the most optimistic situation is that the FST and FCT spectra of $P^{}_{41}$, $P^{}_{42}$ and $P^{}_{43}$ can be observed at two or more different baselines, therefore these different measurements can be cross-checked with each other. Nevertheless, the most promising way is to measure the  $\Delta m^{2}_{41}$ with the frequency spectra from the near detector at short baselines (e.g. 500 m or shorter), for the near detectors can provide large statistics as well as require relative low energy resolution. It is worth to mention that the main advantage of the Fourier transformation technique is that one can easily draw the signal of newly introduced oscillatory terms without precise knowledge of the detector antineutrino spectrum. However, to precisely determine the value of $\Delta m^{2}_{41}$, the energy response function of the detectors and the uncertainties of the standard mixing parameters should be further carefully evaluated.

\subsection{On the Neutrino Mass Hierarchy}

\begin{figure}
\begin{center}
\vspace{0cm}
\includegraphics[scale=0.6, angle=0, clip=0]{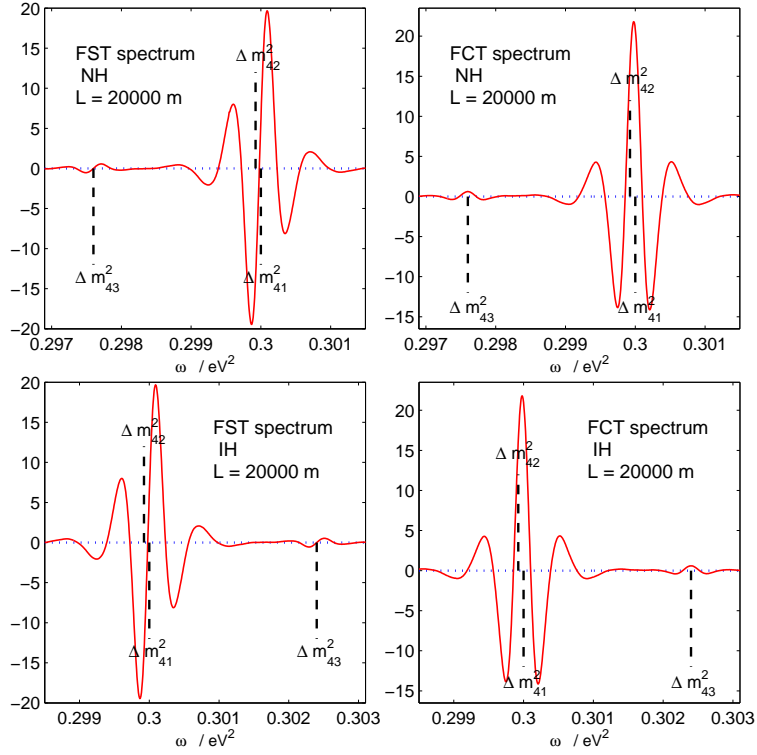}
\vspace{0cm}
\caption{Fourier sine (FST) and cosine (FCT) transformation spectra at the baseline of 20 km for the standard three active neutrinos case (dotted line) and (3+$\mathbbm{1}$) case with $|V^{}_{e4}|^2 = 0.01$ (solid line).}
\end{center}
\end{figure}

As we have mentioned above, the existence of one sub-eV sterile neutrino $\nu^{}_{4}$ will add three new oscillatory components $P^{}_{41}$, $P^{}_{42}$ and $P^{}_{43}$ in $P(\bar{\nu}^{}_{e} \rightarrow \bar{\nu}^{}_{e})$ correspond to three oscillatory frequencies which are proportional to $\Delta^{}_{41}$, $\Delta^{}_{42}$ and $\Delta^{}_{43}$, respectively (with the relative amplitude $c^{2}_{12} c^{2}_{13} : s^{2}_{12} c^{2}_{13} : s^{2}_{13} \approx 28 : 13 : 1$). 
Depending on the mass hierarchy of three active neutrinos (i.e., the sign of $\Delta m^{2}_{31}$), there are two possible ordering of these three new mass-squared differences:
\begin{itemize}
\item Normal hierarchy (NH) with $\Delta m^{2}_{31} > 0$, then we have $\Delta m^{2}_{43} < \Delta m^{2}_{42} < \Delta m^{2}_{41}$;
\item Inverted hierarchy (IH) with $\Delta m^{2}_{31} < 0$, then we have $\Delta m^{2}_{42} < \Delta m^{2}_{41} < \Delta m^{2}_{43}$.
\end{itemize}
It means that the ordering of $\Delta m^{2}_{41}$, $\Delta m^{2}_{42}$ and $\Delta m^{2}_{43}$ is just an indication of the sign of $\Delta m^{2}_{31}$.
Figure 7 shows the frequency spectra of  the three new oscillatory components $P^{}_{41}$, $P^{}_{42}$ and $P^{}_{43}$, in which the main waves are the superposed frequency spectra of $P^{}_{41}$ and $P^{}_{42}$ and the oscillatory term $P^{}_{43}$ modulates the spectra with a small fluctuation at a fixed distance about $2.4 \times 10^{-3}_{} ~ {\rm eV}^2$ away from the main waves. If the frequency spectrum of $P^{}_{43}$ lies at lower frequency than the spectra of $P^{}_{41}$ and $P^{}_{42}$, one can then conclude that $\Delta m^{2}_{31} > 0$. On the contrary, If the spectrum of $P^{}_{43}$ lies at higher frequency than that of $P^{}_{41}$ and $P^{}_{42}$, then we must have $\Delta m^{2}_{31} < 0$. 

Although such a measurement is theoretically feasible, it is in practice challenging. Firstly, a relative long baseline is needed so as the spectrum of $P^{}_{43}$ can be separated from the main spectra of $P^{}_{41}$ and $P^{}_{42}$. We find that the minimum baseline is 20 km for this propose, as shown in Fig. 7. Meanwhile extremely high energy resolution and large statistic are required so as the spectra of these three high-frequency oscillatory terms are not smeared out and the small amplitude fluctuation of $P^{}_{43}$ can be observed. The longer the baseline is, the higher experimental requirements of energy resolution and statistic are required. Therefore, it should be considered only as a complementary to the measurement by the analysis of the frequency spectra of three standard oscillatory terms $P^{}_{21}$, $P^{}_{31}$ and $P^{}_{32}$ \cite{Zhan2008,resolution,MH}.

\section{Summary}

Even though there have been many positive hints of the possible existence of sterile neutrinos and small unitarity violation in the MNSP matrix from both the theoretical and the experimental sides, there is currently no definite constraint on the mass of these particles. It is one of the important jobs to determine or constrain the number of sterile neutrinos and their mass and mixing properties in future precision experiments. The existence of sterile neutrinos can produce various kinds of effects on neutrino oscillations depending on the properties of the sterile neutrinos (e.g., the scale of the sterile neutrino mass, the magnitude and the structure of the active-sterile mixing) as well as the configurations of the experiments (e.g., the oscillation channel, the energy spectrum of the neutrino flux, the baseline $L$, whether the matter effect need to be taken into account). 

In this paper we studied the possibility of searching for sub-eV sterile neutrinos in the precision reactor antineutrino oscillation experiments with three different baselines at around 500 m, 2 km and 60 km respectively. The strategy of placing functionally identical detectors at different baselines and carrying out a combined analysis can offer a ``clean" measurement of the electron antineutrino survival probabilities which is CP-phases independent as well as antineutrino flux independent. We found that the active-sterile mixing angle $\theta^{}_{14}$ could be determined or constrained by the precision measurement of the relative event rate $P(L = 60000 ~ {\rm m})/P(L = 500 ~ {\rm m})$, provided that $\theta^{}_{13}$, $\theta^{}_{12}$ and $\Delta m^{2}_{21}$ were well determined. The mass-squared difference $\Delta m^{2}_{41}$ could be obtained from the Fourier transformation to the $L / E$ spectrum at the near detector.

We underline that the antineutrino survival probabilities obtained in reactor experiments are sensitive only to the direct unitarity violation which is induced by the existence of light sterile neutrinos but independent of the indirect unitarity violation parameters. More specifically, the reactor experiments offer very concentrated sensitivity only to two of the direct unitarity violation parameters $\theta^{}_{14}$ and $\Delta m^{2}_{41}$. This means if any signals of unitarity violation are observed, we can then draw some definite informations on the mass and mixing properties of the light sterile neutrinos. On the contrary, if no observable effect of the unitarity violation are found in the reactor experiments, strong constraints on $\theta^{}_{14}$ and $\Delta m^{2}_{41}$ should be obtained without the possibilities of cancelations between different unitarity violation effects. 

Surely, for such measurements to succeed, both the high energy resolution and the large statistics are essentially important. With the aim of determining the neutrino mass hierarchy, the upcoming JUNO experiment plan to build a 20 kton liquid scintillator detector of the $3\% / \sqrt{E({\rm MeV})}$ energy resolution at about 52 km from reactors of total thermal power 36 GW. To find out the accurate possibility of searching sub-eV sterile neutrinos in this kind of precision reactor experiment, a detailed $\chi^2$ analysis that incorporates available information from experiments and all other uncertainties is need.
Also, accurate informations on the standard mass and mixing parameters are crucial for determining the unitarity violation parameters. The global analysis of various oscillation experiments are highly required for the complete determination of the full mass and mixing pattern of the active and sterile neutrinos \cite{Global}.

\begin{acknowledgements}

I would like to thank Z. Z. Xing and Y. F. Li for helpful discussions. This work was supported in part by the National Natural Science Foundation of China under Grant No. 11105113.

\end{acknowledgements}

\begin{appendix}

\section{On the general ($3$+$\mathbbm{1}$+$\mathbf{N}$) scenario}

In this appendix, we consider a more general ($3$+$\mathbbm{1}$+$\mathbf{N}$) scenario, in which $\mathbbm{1}$ light sterile neutrino $\nu^{}_{s}$ and $\mathbf{N}$ heavy sterile neutrinos $\nu^{}_{h_1}$, $\nu^{}_{h_2}$, ..., $\nu^{}_{h_N}$ are added to the standard three active neutrinos framework. In the ($3$+$\mathbbm{1}$+$\mathbf{N}$) scenario, the full picture of the neutrino mixing should be described by a $n \times n$ unitary matrix $V$
\begin{equation}
\left ( \begin{matrix} \nu^{}_{e} \cr \nu^{}_{\mu} \cr \nu^{}_{\tau} \cr \nu^{}_{s} \cr \nu^{}_{h_1} \cr \vdots \cr \nu^{}_{h_N} \cr \end{matrix} \right ) \; = \; \left ( \begin{matrix} V^{}_{e1} & V^{}_{e2} & V^{}_{e3} & V^{}_{e4} & V^{}_{e5} & \cdots & V^{}_{en} \cr V^{}_{\mu 1} & V^{}_{\mu 2} & V^{}_{\mu 3} & V^{}_{\mu 4} & V^{}_{\mu 5} & \cdots & V^{}_{\mu n} \cr V^{}_{\tau 1} & V^{}_{\tau 2} & V^{}_{\tau 3} & V^{}_{\tau 4} & V^{}_{\tau 5} & \cdots & V^{}_{\tau n} \cr V^{}_{s1} & V^{}_{s2} & V^{}_{s3} & V^{}_{s4} & V^{}_{s5} & \cdots & V^{}_{sn} \cr V^{}_{h_1 1} & V^{}_{h_1 2} & V^{}_{h_1 3} & V^{}_{h_1 4} & V^{}_{h_1 5} & \cdots & V^{}_{h_1 n} \cr \vdots & \vdots & \vdots & \vdots & \vdots & \ddots & \vdots \cr V^{}_{h_N 1} & V^{}_{h_N 2} & V^{}_{h_N 3} & V^{}_{h_N 4} & V^{}_{h_N 5} & \cdots & V^{}_{h_N n} \cr \end{matrix} \right ) \; \left ( \begin{matrix} \nu^{}_{1} \cr \nu^{}_{2} \cr \nu^{}_{3} \cr \nu^{}_{4} \cr \nu^{}_{5} \cr \vdots \cr \nu^{}_{n} \end{matrix} \right ) \; ,
%      (a1)
\end{equation}
where $n = 3+1+N$. In this scenario, there are all together $(n-1)^2$ independent mixing parameters (including $\displaystyle \frac{1}{2}n(n-1)$ mixing angles and $\displaystyle \frac{1}{2}(n-1)(n-2)$ phases) and $n$ neutrino masses.

However, for low energy experiments, no matter how large $N$ is, only the elements in the $3 \times 4$ left-up sub-matrix $U$ of $V$ are related to the neutrino oscillation probabilities:
\begin{eqnarray}
\left ( \begin{matrix} \nu^{}_{e} \cr \nu^{}_{\mu} \cr \nu^{}_{\tau} \end{matrix} \right ) \; = \; U \left ( \begin{matrix} \nu^{}_{1} \cr \nu^{}_{2} \cr \nu^{}_{3} \cr \nu^{}_{4} \end{matrix} \right ) \; = \; \left ( \begin{matrix} V^{}_{e1} & V^{}_{e2} & V^{}_{e3} & V^{}_{e4} \cr V^{}_{\mu 1} & V^{}_{\mu 2} & V^{}_{\mu 3} & V^{}_{\mu 4} \cr V^{}_{\tau 1} & V^{}_{\tau 2} & V^{}_{\tau 3} & V^{}_{\tau 4} \end{matrix} \right ) \; \left ( \begin{matrix} \nu^{}_{1} \cr \nu^{}_{2} \cr \nu^{}_{3} \cr \nu^{}_{4} \end{matrix} \right ) \; .
%      (a2)
\end{eqnarray}
Here $U$ is in general non-unitary and consist of at most 24 independent real parameters (some unphysical phases are also counted). Which means for any $N \geq 2$, neutrino oscillation probabilities for any low energy experiments can all be effectively parametrized by 24 independent mixing parameters and 3 independent mass-squared differences.  In the special case of $N = 1$, $U$ can be parametrised by only 16 independent mixing parameters (10 mixing angles and 6 phases) as we have explained in Sec. I.

\end{appendix}

\end{document}